\documentclass[fleqn]{2023SCGE}
\begin{document}

\ensubject{subject}

%%%%%%%%%%%%%%%%%%%%%%%%%%%%%%%%%%%%%%%%%%%%%%%%%%%%%%%
%%% Authors do not modify the information below
%%% ????????????????
%%% ??????????, ????????????{}, ???????????????????
%Letter to the Editor??Article%??????
\ArticleType{Article}%??Article
%\SpecialTopic{SPECIAL TOPIC: }%???????
\Year{0000}
\Month{00}
\Vol{00}
\No{0}
\DOI{??}
\ArtNo{000000}
\ReceiveDate{xxx 00, 0000}
\AcceptDate{xxx 00, 0000}
%\OnlineDate{January 1, 2016}
%%%%%%%%%%%%%%%%%%%%%%%%%%%%%%%%%%%%%%%%%%%%%%%%%%%%%%%

%%% title: ????
%%%   \title{title}{title for citation}
\title{Testing the first law of black hole mechanics with gravitational waves}

%%% Corresponding author: ???????
%%%   \author[number]{Full name}{{email@xxx.com}}
%%% General author: ???????
%%%   \author[number]{Full name}{}
\author[1,2]{Chao-Wan-Zhen Wang}{}
\author[1,2]{Jin-Bao Zhu}{}
\author[1,2]{Guo-Qing Huang}{}%\protect\\?��??��????
\author[1,2,3]{Fu-Wen Shu}{email@shufuwen@ncu.edu.cn}%
%\author[1]{E. Author}{}

%%% Author information for page head. ?��?��????????
%%% ??????????????, ??????????author???
\AuthorMark{C.-W.-Z. Wang}%\authorcr????????

%%% Authors for citation. ????????��????????
%%% ??????????????, ??????????author???
\AuthorCitation{C.-W.-Z. Wang, J.-B. Zhu, G.-Q. Huang and F.-W. Shu}

%%% Address. ???
%%%   \address[number]{Address, City {\rm Postcode}, Country}
\address[1]{Department of Physics, Nanchang University, Nanchang, 330031, China}
\address[2]{Center for Relativistic Astrophysics and High Energy Physics, Nanchang University, Nanchang 330031, China}
\address[3]{GCAP-CASPER, Physics Department, Baylor University, Waco, Texas 76798-7316, USA}

%\contributions{}%????????

%%% Abstract. ??
\abstract{The successful observation of gravitational waves has provided humanity with an additional method to explore the universe, particularly black holes. In this study, we utilize data from LIGO and Virgo gravitational wave observations to test the first law of black hole mechanics, employing two different approaches. We consider the secondary compact object as a perturbation to the primary black hole before the merger, and the remnant black hole as a stationary black hole after the merger. In the pre-merger and post-merger analysis, our results demonstrate consistency with the first law, with an error level of approximate 25\% at a 68\% credibility level for GW190403\_051519. In the full inspiral-merger-ringdown analysis, our results show consistency with the first law of black hole mechanics, with an error level of about 6\% at a 68\% credibility level and 10\% at a 95\% credibility level for GW191219\_163120. Additionally, we observe that the higher the mass ratio of the gravitational wave source, the more consistent our results are with the first law of black hole mechanics. Overall, our study sheds light on the nature of compact binary coalescence and their implications for black hole mechanics.}%ժҪ

%%% Keywords. ?????


\maketitle

%\tableofcontents%?????

%%%%%%%%%%%%%%%%%%%%%%%%%%%%%%%%%%%%%%%%%%%%%%%%%%%%%%%
%%% The main text. ???????
%???????????????????\cref{fig1}
%\twocolumn\onecolumn
%%%%%%%%%%%%%%%%%%%%%%%%%%%%%%%%%%%%%%%%%%%%%%%%%%%%%%%
\begin{multicols}{2}%2????????

\section{Introduction}\label{sec:1}
Black holes (BHs)  are among the most fascinating objects in the universe, with their fruitful field for theoretical exploration of the frontiers of physics, including topics such as the singularity theorem, the cosmic censorship conjecture, and the black hole information paradox \cite{Penrose:1964wq,Penrose:1969pc,Hawking:1976ra}. Over the past few decades, our understanding of BHs has improved significantly, particularly in the areas of gravitation, thermodynamics, and quantum theory. BH mechanics, in particular, plays a crucial role in the relationship between these fields and has provided us with a deeper understanding of quantum phenomena that occur in strong gravitational fields \cite{wald2001thermodynamics}.

Testing BH mechanics has become increasingly important for developing a comprehensive theory of quantum gravity. The advent of gravitational wave (GW) astronomy has opened a new window on the study of BHs \cite{newwindow,astronomy}. In 2015, the LIGO detectors observed a GW signal from a binary BH system, GW150914, which was the first direct detection of GWs by human beings \cite{abbott2016observation}. Since then, a number of GW events have been observed, providing valuable insights into BH mergers and their properties \cite{review,review2}. In addition, observational testing of BH properties is made possible with the aid of various methods. For example, the black-hole area law, also known as the second law of BH mechanics, has been tested \cite{isi2021testing}, along with the BH no-hair theorem \cite{isi2019testing}. Here, we present a new GW observational test on the first law of BH mechanics.

The first law of BH mechanics, also known as the Bekenstein-Smarr formula, has important implications for our understanding of BHs and their properties. This law describes an important relationship between the parameters of a stationary axisymmetric BH if the BH settles down to a new stationary axisymmetric BH after some infinitesimal physical process \cite{PhysRevD.7.2333,PhysRevLett.30.71,bardeen1973four,wald1994quantum}. 
While the second law of BH mechanics has been tested with GWs \cite{isi2021testing}, the first law has yet to be confirmed.

GWs carry valuable information about their sources. By analyzing the data from GW observations, we can indirectly obtain information about the sources. In this paper, we propose a scheme to test the first law of BH mechanics using GWs that have been observed so far. We approach this testing from two perspectives: the pre-merger and post-merger analysis, and the full inspiral-merger-ringdown (IMR) analysis. Our results provide important insights into the testing of BH mechanics and its implications for the development of a comprehensive theory of quantum gravity.

In order to view the merger of a binary system as a perturbed process and test the first law of BH mechanics, it is essential to ensure that the chosen GW event has a large mass ratio. When the collision occurs, we consider the secondary compact object as a perturbation to the primary black hole. We conducted separate analyses using the pre-merger and post-merger data of GW190403\_051519, with mass ratios of approximately 4. Additionally, we analyzed the complete IMR data of all GW events with a mass ratio higher than 3 that have been observed thus far. Among these events, GW191219\_163120 exhibits the highest mass ratio of 27 and serves as an excellent candidate for our analysis. 
\section{Method}\label{sec:2}
The first law of BH mechanics is a fundamental result in the study of BHs that relates changes in the mass, angular momentum, and area of a stationary BH when it perturbs. It states that for a stationary BH in a vacuum, the variations of these quantities satisfy \cite{PhysRevD.7.2333,PhysRevLett.30.71,bardeen1973four,wald1994quantum}:
\begin{equation}\label{firstlaw}
	\delta M = \frac{\kappa}{8\pi}\delta A + \Omega \delta J,
\end{equation}
where $M$, $J$, and $A$ are the mass, angular momentum, and area of the BH, respectively. $\kappa$ is the surface gravity of the BH and $\Omega$ is its angular velocity at the event horizon. In terms of the mass, $M$, and the dimensionless spin magnitude, $ \chi \equiv |\vec{J}| c/(GM^2)$, these quantities have the following explicit form:

\begin{align}
	\begin{array}{l}
		\kappa = \dfrac{c^{4}\sqrt{1-\chi^{2}}}{G^{2}M(1+\sqrt{1-\chi ^{2}})},\\[12pt]
	    A =  \dfrac{8\pi G^2M^2}{c^{4}}\left(1+\sqrt{1-\chi^{2}}\right),\\[8pt]
	    \Omega = \dfrac{cM\chi}{2G(r_{+}^{2}+M^{2}\chi^{2})},
	\end{array}
\end{align}
where $r_{+}$ is radius of the outer horizon and satifies
\begin{equation}
	\begin{array}{l}
		r_{+} = M\left(1+\sqrt{1-\chi^{2}}\right).
	\end{array}
\end{equation}

Eq. \eqref{firstlaw} is analogous to the first law of thermodynamics, which relates changes in energy, heat, and work as $\delta E = T\delta S - P\delta V$. The first term on the right-hand side of the equation, $\frac{\kappa}{8\pi}\delta A$, represents the change in the BH's energy due to a change in its area. The second term, $\Omega \delta J$, represents the work done on the BH by a change in its angular momentum \cite{PhysRevD.7.2333}. As a consequence, the first law of BH mechanics is an important result in the study of BHs, as it suggests that BHs possess properties similar to those of thermodynamic systems, such as temperature and entropy. In addition, a remnant BH after binary black hole merger is thermodynamically stable \cite{thermodynamicallystable}.

To assess the validity of the first law of BH mechanics, we define $r$ as $r = \delta M/\left(\frac{\kappa}{8\pi}\delta A + \Omega \delta J\right)$ to calculate the ratio between the left and right-hand sides of Equation \eqref{firstlaw}. By determining the probability that the value of $r$ approaches unity, we can evaluate the extent to which the first law holds. The probability density distribution (PDD) of ratio $r$ can be obtained by combining the PDD of the following quantities: $\delta M = m_{f}-m_{0}$, $\delta A=A_f-A_0$, $\delta J=J_f-J_0$, $\kappa$ and $\Omega$. Here the subscripts ``0'' and ``$f$'' represent the quantities before and after the merger of the primary BH, respectively.

To obtain the PDDs of $A_{0,f}$ and $J_{0,f}$ for BHs before and after mergers, we can combine the posterior distributions of their masses $m_{0,f}$ and dimensionless spin magnitudes $\chi_{0,f}$, as described in \cite{cabero2018observational}. Similarly, the PDDs of the surface gravity $\kappa$ and orbital frequency $\Omega$ before the merger can be obtained by combining the posterior PDDs of the initial primary BH mass $m_{0}$ and spin magnitude $\chi_{0}$. The PDDs of the differences in mass $\delta M$, in horizon area $\delta A$, and in angular momentum $\delta J$ then can be obtained by combining the PDDs of $m_{0,f}$, $A_{0,f}$, and $J_{0,f}$. 

In order to evaluate the reliability of our GW observed data, we need to compare the actual value of $r$ with the theoretical expected value. We can quantify the accuracy of our results using the fractional difference, $ (r_{a} - r_{e})/r_{e}$, where $r_{a}$ represents the actual value of $r$ and $r_{e}$ represents the expected value of $r$, which is 1 \cite{scientific2016tests}.

By calculating this fractional difference, we can measure the level of error in our data and determine how closely it matches our theoretical expectations. This is an important step in ensuring the accuracy and validity of our results, and can help us to draw more meaningful conclusions about the nature of GWs.

The observed data in gravitational wave (GW) analysis consists of both signals and noise. The GW signals are embedded within the data, but their detectable strain in the space-time fabric is typically around $10^{-19}$, whereas the data itself exhibits a strain on the order of $10^{-21}$. To achieve more accurate parameter estimation for GW sources, it is necessary to preprocess the data by applying whitening and bandpass filtering techniques prior to template matching.

In the context of parameter estimation for gravitational wave (GW) applications, Bayes's theorem can be represented as follows:
\begin{equation}\label{bayes}
P(\theta,M|d)=\dfrac{L(d|\theta,M)P(\theta,M)}{Z},
\end{equation}
where $Z=\int L(d|\theta,M)P(\theta,M)d\theta$ represents the evidence. In this equation, $\theta, M$ and $d$ represent, respectively, the unknown parameters, the waveform templates, and the data. The conditional probability density function of the unknown parameters, given the waveform templates and the data, is denoted as $P(\theta,M|d).$

Obtaining the PDD of parameters often requires the application of computational techniques, such as Markov chain Monte Carlo (MCMC) or nested sampling. These techniques are particularly useful for solving problems involving high-dimensional integration and a large number of iterations, often reaching tens of thousands. In this iterative process, the posterior value from the previous iteration serves as the prior value for the current iteration. The collection of posterior values after convergence constitutes the posterior PDD of the parameters.

In our study, we conduct separate analyses on the pre-merger and post-merger data. For the pre-merger data, we estimate $m_{0}$ and $\chi_{0}$ utilizing the Bilby\footnote{https://lscsoft.docs.ligo.org/bilby/.} \cite{bilby} package in the frequency domain. For the post-merger data, we estimate $m_{f}$ and $\chi_{f}$ utilizing pyRing\footnote{https://lscsoft.docs.ligo.org/pyring/} \cite{pyRing} Python package in the time domain. The pyRing package can measure remnant parameters by analyzing ringdown data directly and independently.

The posterior samplers of $m_{0,f}$ and $\chi_{0,f}$ in our full inspiral-merger-ringdown analysis were obtained from LIGO and Virgo released\footnote{https://zenodo.org/record/6513631/.}\footnote{https://zenodo.org/record/5546663/.}, where $m_{f}$ and $\chi_{f}$ were derived using numerical-relativity fits. The main distinction between the pre-merger and post-merger analysis, and the full inspiral-merger-ringdown analysis, lies in the approach employed to determine $m_{f}$ and $\chi_{f}$. The former analysis directly and independently measures the remnant parameters without assuming the validity of general relativity (GR), while the latter method assumes the validity of GR a priori.

Our estimation of $m_{f}$ and $\chi_{f}$ relies on the ringdown waveform, which can be expressed as:
\begin{equation}\label{strain}
h_{+}(t)-ih_{\times}(t)=\frac{M_{f}}{D_{L}}\sum_{l=2}^{\infty}\sum_{m=-l}^{l}\sum_{n=0}^{\infty}A_{lmn}S_{lmn}(\iota,\varphi)\Psi_{lmn},
\end{equation}
where $\Psi_{lmn}=\exp[(t-t_{lmn})\omega_{lmn}+\phi_{lmn}]$. The strains $h_{+}$ and $h_{\times}$ correspond to the plus and cross polarization of gravitational waves (GWs), respectively. $M_{f}$ represents the final remnant mass, while $D_{L}$ denotes the luminosity distance. The triplet $(l,m,n)$ serves as the three indices labeling the quasi-normal modes (QNMs) of a Kerr black hole, where $l$ and $m$ are respectively the angular and magnetic quantum numbers, and $n$ is the overtone index. The quantities $A_{lmn}$, $t_{lmn}$, $\omega_{lmn}$, and $\phi_{lmn}$ are complex amplitudes, starting times, complex frequencies, and initial phases associated with each mode. $S_{lmn}$ represents the spin-weighted spheroidal harmonics, while $\iota$ and $\varphi$ represent the polar and azimuthal angles, respectively. In our study, we specifically focus on the fundamental mode (with $n=0$) of the dominant mode (with $l=m=2$).

\section{The pre-merger and post-merger analysis}\label{sec:3}

In this section, we analyze the GW event GW190403\_051519 \cite{GW190403}, which has a mass ratio of approximately 4.
This event corresponds to a binary black hole coalescence signal observed during the first half of the third observation run of Advanced LIGO and Advanced Virgo. 

We employ the Bilby package along with the Dynasty sampler to estimate the values of $m_{0}$ and $a_{0}$. The summary of the parameters for GW190403\_051519 estimated with Billy is showed in Table \ref{tableA3} in the Appendix. The waveform model utilized is IMRPhenomXPHM \cite{IMRPhenomXPHM}. The priors for the mass ratio, chirp mass, and $\chi$ are set as [0.1, 1], [50, 100], and [0, 1] respectively. For the GW190403\_051519 event, we determine the values of $m_{f}$ and $a_{f}$ using the pyRing package and the CPnest sampler. The summary of the parameters for GW190403\_051519 estimated with pyRing is showed in Table \ref{tableA4} in the Appendix. The priors for $m_{f}$, $\chi_{f}$, and the starting time of the ringdown are chosen as [140, 310], [0.6, 1], and 1238303737.22 respectively. All masses are measured in the detector frame and are expressed in solar masses.
The specific values of $m_{0,f}$ and $a_{0,f}$ are presented in Table \ref{table1} additionally.

We present the result of testing the first law of the BH mechanics with GW190403\_051519 in Fig.\ref{figure1}. The 68\% credible region is (-25.02\%, 1.33\%), yielding an approximate 25\% error level.

\begin{table*}[t]
	\tabcolsep 13pt
	\centering
	\caption{The parameter values are obtained from different analyses. $m_0^\mathrm{pre}$ and $\chi_0^\mathrm{pre}$ are measured from the pre-merger data, $m_\mathrm{f}^\mathrm{post}$ and $\chi_\mathrm{f}^\mathrm{post}$ are measured from the post-merger date, $m_\mathrm{f}^\mathrm{IMR}$ and $\chi_\mathrm{f}^\mathrm{IMR}$ are measured from numerical-relativity fits. The median value and the range of the 90\% credible interval are provided. All masses are measured in the detector frame and are expressed in solar masses.}
	\label{table1}
	\begin{tabular*}{\textwidth}{ccccccc}
		\toprule
		Event& $m_0^\mathrm{pre}$& $m_\mathrm{f}^\mathrm{post}$ & $m_\mathrm{f}^\mathrm{IMR}$ & $\chi_0^\mathrm{pre}$ & $\chi_\mathrm{f}^\mathrm{post}$ & $\chi_\mathrm{f}^\mathrm{IMR}$ \\
		\hline
		GW190403\_051519 & $182.81_{-76.86}^{+40.47}$& $206.63_{-48.73}^{+73.01}$ & $225.47_{-47.45}^{+36.53}$  & $0.83_{-0.70}^{+0.14}$ & $0.84_{-0.21}^{+0.14}$& $0.91_{-0.17}^{+0.05}$  \\
		\bottomrule
	\end{tabular*}
\end{table*}

\newlength{\multicolwidth}
\setlength{\multicolwidth}{0.5\textwidth}
\addtolength{\multicolwidth}{-0.5\columnsep}
\begin{figure*}[t]
\begin{minipage}[b]{\multicolwidth}
	\centering
	\includegraphics[width=0.9\linewidth]{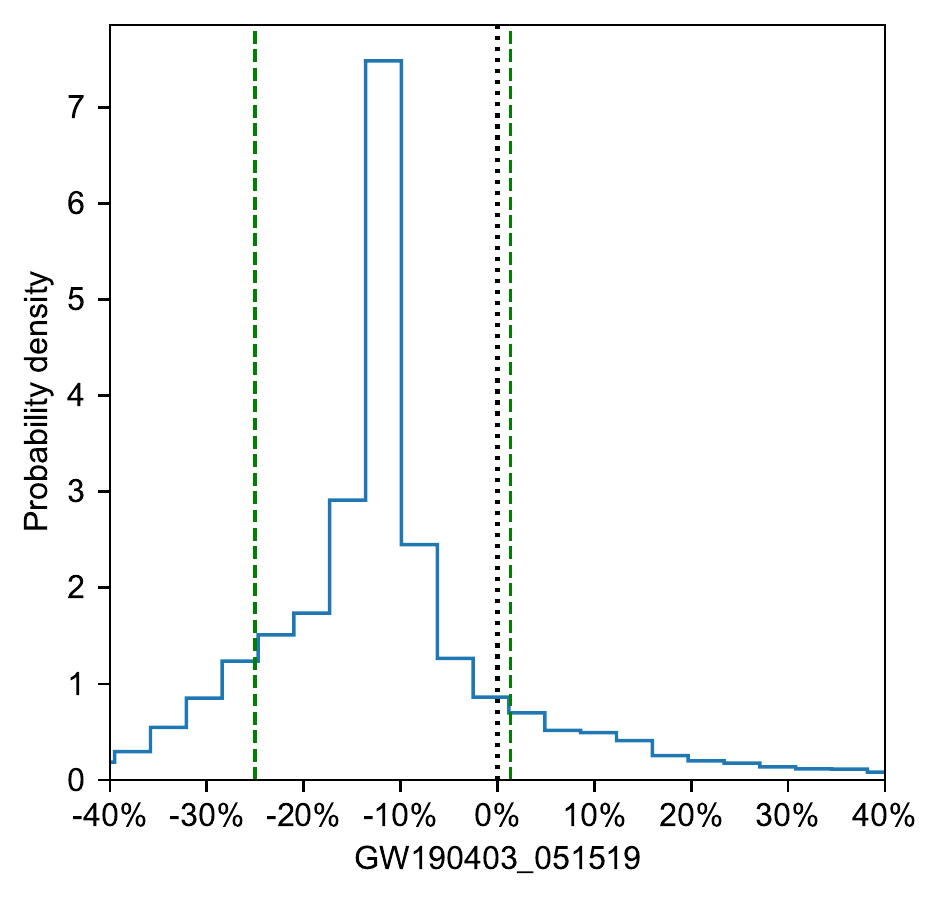}
\end{minipage}\hfill
\begin{minipage}[b]{\multicolwidth}
	\centering
	\includegraphics[width=\linewidth]{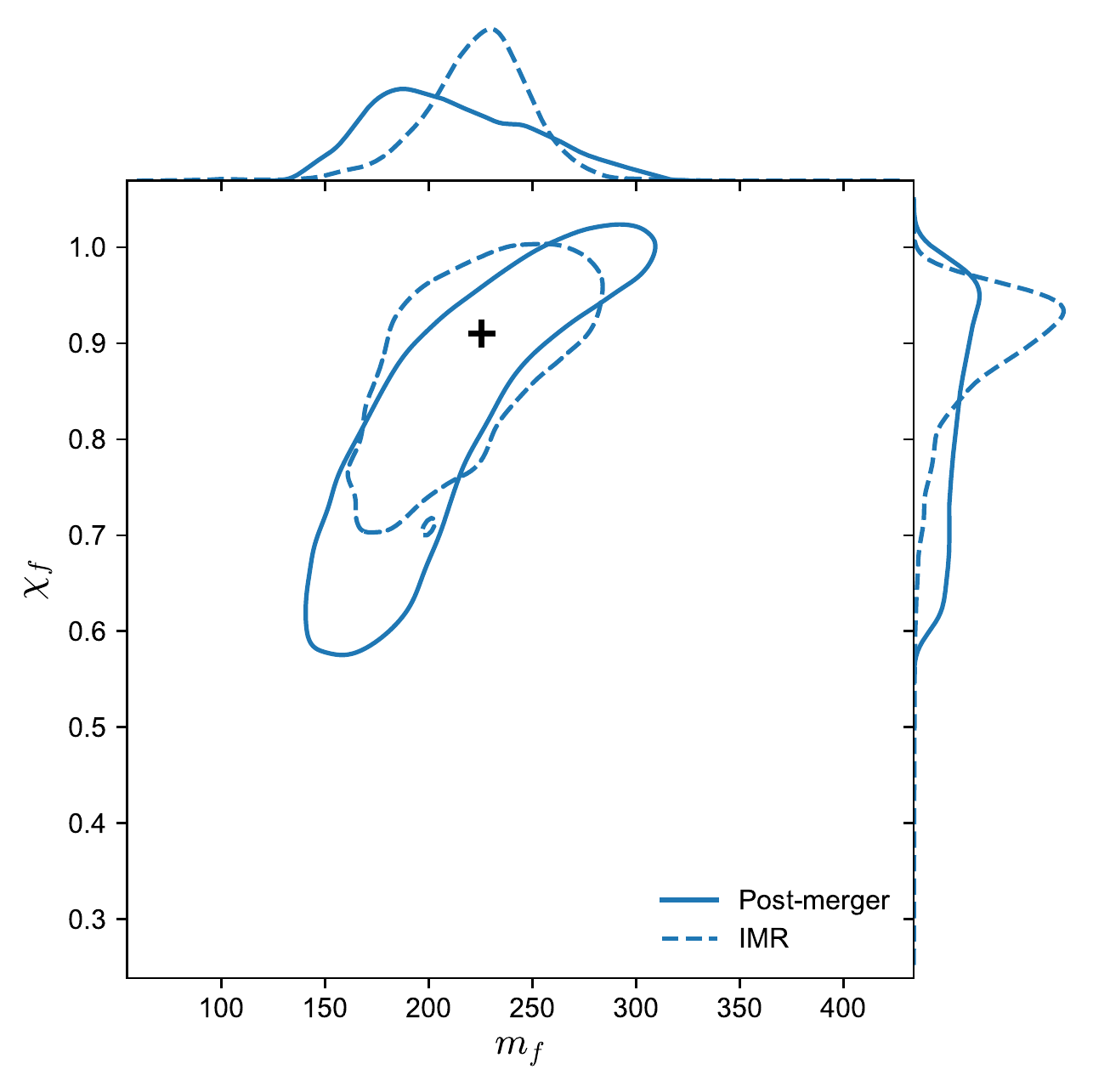}
\end{minipage}\\[-4ex]
\begin{minipage}[t]{\multicolwidth}
	\caption{The probability density distribution of the fractional difference of $r$ of GW190403\_051519, whose mass ratio is about 4. We estimate $m_{0}$ and $\chi_{0}$ with the Bilby package and $m_{f}$ and $\chi_{f}$ with the pyRing package. The green dashed lines indicate the 68\% credible regions. The black dotted line indicates the expected value, which is 0.}
	\label{figure1}
\end{minipage}\hfill
\begin{minipage}[t]{\multicolwidth}
	\caption{The remnant parameters of GW190403\_051519 are measured from the post-merger data analysis (solid blue) and the full IMR analysis (dashed blue). The top and right-hand panels display one-dimensional posteriors for $m_{f}$ and $\chi_{f}$, respectively. Contours represent the 90\% credible regions for $m_{f}$ and $\chi_{f}$. The black plus marker indicates the median values of $m_{f}$ and $\chi_{f}$ obtained from the full IMR analysis. Masses are measured in the detector frame and are expressed in solar masses.}
	\label{figure2}
\end{minipage}
\end{figure*}

The predicted values through numerical relativity are denoted as $m_\mathrm{f}^\mathrm{IMR}$ and $\chi_\mathrm{f}^\mathrm{IMR}$, which are obtained through the mass and spin of the initial primary BH and secondary compact object. While the measured values from the data are represented by $m_\mathrm{f}^\mathrm{post}$ and $\chi_\mathrm{f}^\mathrm{post}$, which are derived directly and independently from the analysis of ringdown data. By assessing the consistency between these values, we can test the validity of GR. The values of $m_\mathrm{f}^\mathrm{IMR}$ and $\chi_\mathrm{f}^\mathrm{IMR}$ of GW190403\_051519 are obtained from LIGO and Virgo released\footnote{https://zenodo.org/record/6513631/.}. The results are shown in Fig. \ref{figure2}. As depicted in Fig. \ref{figure2}, 88\% of the samples within the blue dashed contour lie within the overlapping region of the blue solid contour. This indicates an agreement between the measured remnant parameters from both approaches, with an 88\% probability. This result supports the validity of general relativity, and confirms that the remnant black hole of GW190403\_051519 is a Kerr black hole, as predicted by GR.

% \begin{figure}[H]
% 	\centering
% 	\includegraphics[width=0.4\linewidth]{GR.pdf}
% 	\caption{The remnant parameters of GW190403\_051519 are measured from the post-merger data analysis (solid blue) and the full IMR analysis (dashed blue). The top and right-hand panels display one-dimensional posteriors for $m_{f}$ and $\chi_{f}$, respectively. Contours represent the 90\% credible regions for $m_{f}$ and $\chi_{f}$. The black plus marker indicates the median values of $m_{f}$ and $\chi_{f}$ obtained from the full IMR analysis. \textcolor{blue}{Masses are measured in the detector frame and are expressed in solar masses.}}
% 	\label{GR}
% \end{figure}

\section{The full inspiral-merger-ringdown analysis}\label{sec:4}

In the analysis conducted in the previous section, the results of testing the first law were not satisfactory. Our goal is to analyze events with higher mass ratios. However, as the mass ratio increases, the disturbance becomes smaller. Since ringdown signals are already weak, analyzing them becomes even more challenging. Therefore, we attempt a full IMR analysis, where the remnant parameters are obtained through numerical-relativity fits using the pre-merger parameters as raw data, rather than directly analyzing the ringdown signal.

We begin by analyzing GW191219\_163120, which is currently the event with the highest mass ratio observed. GW191219\_163120 was observed during the second part of  Advanced LIGO's and Advanced Virgo's third observing run \cite{abbott2021gwtc}. This event features a BH primary component with a mass of $31.1^{+2.2}_{-2.7}$ and a neutron star secondary component with a mass of $1.17^{+0.07}_{-0.06}$ (at 90\% credible regions). After the merger, the final remnant mass is $32.2^{+2.2}_{-2.7}$. With a mass ratio of 27, which is comparatively extreme, this event provides an opportunity to test the first law of BH mechanics, which pertains to small change.

\begin{figure*}[t]
\begin{minipage}[b]{\multicolwidth}
	\centering
	\includegraphics[width=0.87\linewidth]{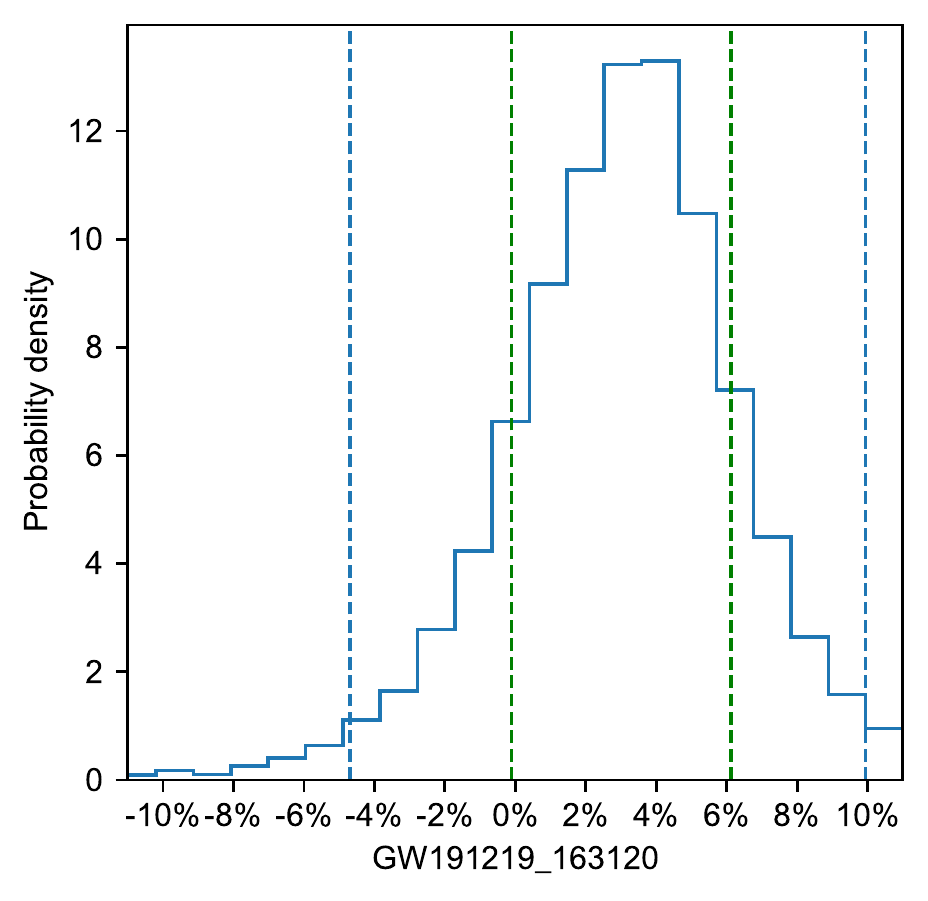}
\end{minipage}\hfill
\begin{minipage}[b]{\multicolwidth}
	\centering
	\includegraphics[width=0.9\linewidth]{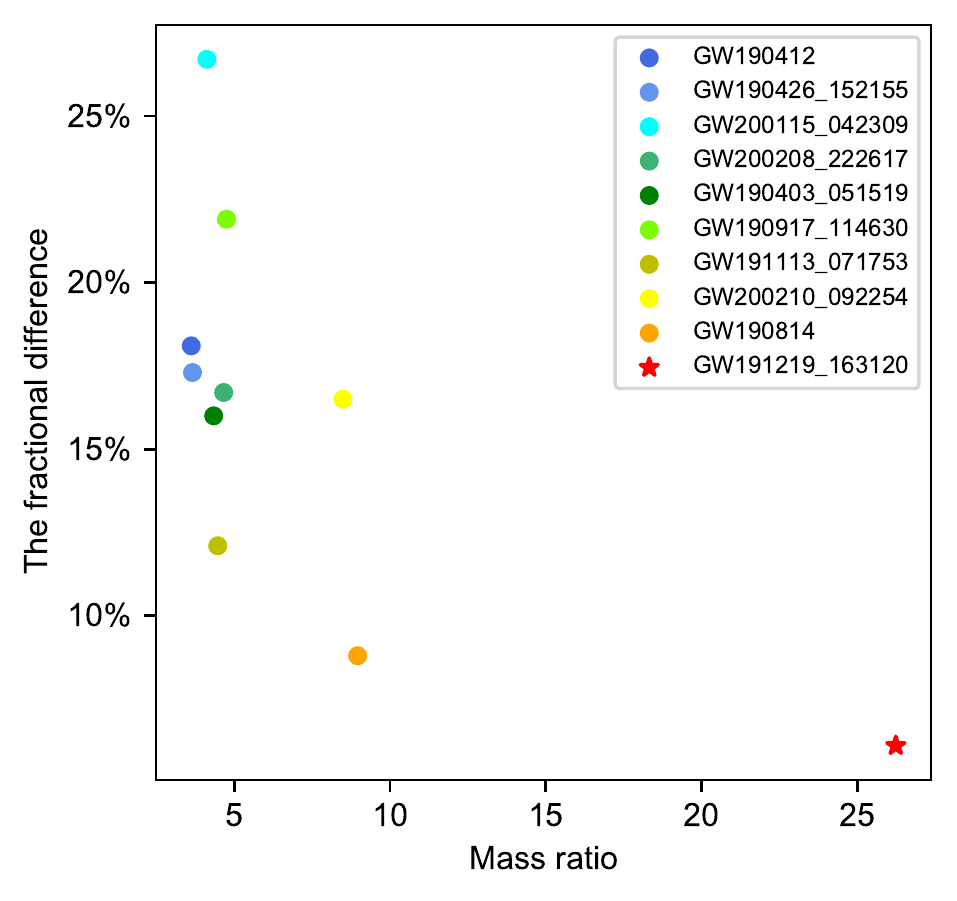}
\end{minipage}\\[-4ex]
\begin{minipage}[t]{\multicolwidth}
	\caption{The probability density distribution of the fractional difference of $r$ is shown, where we mix the samples from the last three waveform models and use 70 bins. The green dashed lines indicate the 68\% credible regions, and the blue dashed lines indicate the 95\% credible regions. The black dotted line indicates the expected value, which is 0.}
	\label{figure3}
\end{minipage}\hfill
\begin{minipage}[t]{\multicolwidth}
	\caption{68\% credible measurements of the fractional difference of different mass ratios of  GW sources. Each point represents a different GW event.}
	\label{figure4}
\end{minipage}
\end{figure*}

The parameter samples for GW191219\_163120 were derived by matching different waveform models. The parameters values vary among different waveform models. To eliminate the effect of waveform models on our results, we combine three waveform models, IMRPhenomXPHM: HighSpin, IMRPhenomXPHM: LowSpin, and SEOBNRv4PHM, with a weight of 1:1:2 to ensure that the weight of IMRPhenom and SEOBNR is 1:1. This is illustrated in Fig.\ref{figure3}. The 68\% credible region is (-0.10\%, 6.13\%), and the 95\% credible region is (-4.70\%, 9.94\%). The result indicates that we have tested the first law of BH mechanics with an error level of about 6\% (10\%) and 68\% (95\%) credibility under the assumption of GR being valid.

The first law of BH mechanics, as shown in Eq. \eqref{firstlaw}, pertains to small variations of $M$, $A$, and $J$. To test this law, we can approximate the merger of a binary system as a perturbed process and examine the fractional difference of $r$. We hypothesize that there should be a negative correlation between the mass ratio of GW sources and the error level. In other words, as the mass ratio of the binary increases, the fractional difference of $r$ should decrease. To verify our hypothesis, we analyze all GW events with a mass ratio higher than 3 that have been observed thus far. Ten events in total meet this criterion, with different mass ratios. Five of these events are from the O3a observation run of Advanced LIGO and Advanced Virgo \cite{abbott2021gwtc1}, specifically GW190403\_051519, GW190412,  GW190426\_152155, GW190814, and GW190917\_114630. The remaining five events are from the O3b observation run \cite{abbott2021gwtc}, specifically GW191113\_071753, GW191219\_163120,  GW200115\_042309, GW200208\_222617, and GW200210\_ 092254.  Our analysis confirms our hypothesis, as demonstrated in Fig. \ref{figure4}. We show the detailed mass ratios and the fractional differences in Table \ref{table2}. The figure shows that, in general, the higher the mass ratio of the GW source, the smaller the fractional difference and therefore, the lower the error level.

\begin{table}[H]
	\tabcolsep 10pt
	\caption{Mass ratios and the fractional differences at 68\% credible regions of twelve events in Fig. \ref{figure4}. The median value and the range of the 90\% credible interval of mass ratios are provided.}
	\label{table2}
	\begin{tabular*}{\columnwidth}{c c c}
		\toprule
		Event & Mass ratio & \parbox[c][9.6mm]{2.4cm}{\centering The fractional\\difference(\%)}\\
		\hline
		GW190412\_053044 & $3.07_{-1.06}^{+1.31}$  & 19.63\\
		GW190426\_152155 & $3.66_{-2.07}^{+4.13}$  & 17.26\\
		GW200115\_042309 & $4.12_{-2.64}^{+2.76}$  & 26.73\\
		GW190403\_051519 & $4.34_{-2.84}^{+4.33}$  & 16.04\\
		GW200208\_222617 & $4.66_{-3.53}^{+13.03}$ & 16.71\\
		GW200105\_162426 & $4.74_{-1.07}^{+1.70}$  & 12.12\\
		GW191113\_071753 & $4.95_{-3.51}^{+3.73}$  & 13.36\\
		GW200210\_092254 & $8.49_{-2.47}^{+4.52}$  & 16.51\\
		GW190814         & $8.96_{-0.62}^{+0.75}$  & 8.79\\
		GW191219\_163120 & $26.65_{-3.29}^{+2.86}$ & 6.13\\
		\bottomrule
	\end{tabular*}
\end{table}

\section{Discussion and conclusions}\label{sec:5}
Our use of GW data observed by LIGO and Virgo provides both physical evidence and an observational test of this mathematically derived law. We are the first to test this law using GWs, which adds to the growing body of knowledge about BHs and their behavior.

Firstly, we test the first law with GW190403\_051519. However, the collision between the primary black hole and secondary black hole does not satisfy the perturbation condition required by the first law, mainly due to the insufficiently high mass ratio of the event. As a result, the constraints for testing the first law are not strong. Unfortunately, we are unable to obtain effective posterior samplers for $m_{f}$ and $a_{f}$ for all events. The ringdown signal is weaker than the pre-merger signal, making the analysis of the ringdown data more challenging.

Therefore, we have turned to an alternative analysis approach. The full IMR analysis can demonstrate that if our analysis of the ringdown signal is sufficiently advanced and refined, the error level of testing the first law will be significantly reduced. In our analysis, GW191219\_163120, which has the highest mass ratio among the events considered, exhibits an error level of approximately 10\% at a 95\% credibility level.

We totally analyze ten events with the full IMR analysis. We find that the error level of testing the first law is related to the mass ratio of the gravitational wave sources, with higher mass ratios resulting in lower error levels and better agreement with the first law of BH mechanics. In the future, ground-based and space GW observatories will be able to further test the first law with even lower error levels and higher agreement with higher mass ratio signals.

Ground-based GW observatories, such as LIGO \cite{martynov2016sensitivity}, Virgo \cite{bersanetti2021advanced}, and KAGRA \cite{kagra2019kagra} , operate in the frequency band of 10Hz$\sim$10kHz, while space GW observatories, such as LISA \cite{shaddock2008space}, Taiji \cite{hu2017taiji,ruan2020taiji,luo2020brief}, and TianQin \cite{luo2016tianqin,Gong:2021gvw} , operate in the frequency band of 0.1 mHz$\sim$1 Hz. Space observatories are more sensitive to low-frequency signals than ground-based observatories and can detect GW signals emitted by a stellar BH orbiting a supermassive BH, known as Extreme Mass-Ratio Inspirals (EMRIs) and Extreme Mass-Ratio Bursts (EMRBs), with a mass ratio of at least $10^{4}$ \cite{shaddock2008space}. EMRI signals are continuous due to their low eccentric orbits \cite{amaro2018relativistic}, while EMRB signals are short-lived due to their high eccentric orbits \cite{berry2013observing}. LISA, for example, can detect a minimum of a few EMRIs events per year and a maximum of a few thousand EMRIs events per year \cite{babak2017science}, and two Galactic EMRBs events could be observed in LISA's two-year lifetime \cite{berry2013expectations}. By analyzing events with a mass ratio of at least $10^{4}$, the error level of testing the first law of BH mechanics will be significantly lower, and the agreement with the first law will be higher.

%%%%%%%%%%%%%%%%%%%%%%%%%%%%%%%%%%%%%%%%%%%%%%%%%%%%%%%
%%% Acknowledgements. ??��
%%%%%%%%%%%%%%%%%%%%%%%%%%%%%%%%%%%%%%%%%%%%%%%%%%%%%%%
\Acknowledgements{This study is supported by the National Natural Science Foundation of China with Grant Nos.  12375049, 11975116, and Key Program of the Natural Science Foundation of Jiangxi Province under Grant No. 20232ACB201008.  This research has made use of data
	obtained from the Gravitational Wave Open Science Center\footnote{https://www.gw-openscience.org/.}, a service of LIGO Laboratory, the LIGO Scientific Collaboration and the Virgo Collaboration. LIGO is funded by the U.S. National Science Foundation. Virgo is funded by the French Centre National de Recherche Scientifique (CNRS), the Italian Istituto Nazionale della Fisica Nucleare (INFN), and the Dutch Nikhef, with contributions by Polish and Hungarian institutes.}

%%%%%%%%%%%%%%%%%%%%%%%%%%%%%%%%%%%%%%%%%%%%%%%%%%%%%%%
%%% Conflict of interest. ????????????
%%%%%%%%%%%%%%%%%%%%%%%%%%%%%%%%%%%%%%%%%%%%%%%%%%%%%%%
\InterestConflict{The authors declare that they have no conflict of interest.}

%%%%%%%%%%%%%%%%%%%%%%%%%%%%%%%%%%%%%%%%%%%%%%%%%%%%%%%
%%% Supplements. ????????, ????
%%%%%%%%%%%%%%%%%%%%%%%%%%%%%%%%%%%%%%%%%%%%%%%%%%%%%%%
%\Supplements{}

%%%%%%%%%%%%%%%%%%%%%%%%%%%%%%%%%%%%%%%%%%%%%%%%%%%%%%%
%%% Reference section. ?��?????
%%% citation in the content using "some words~\cite{1,2}".
%%% ~ is needed to make the reference number is on the same line with the word before it.
%%%%%%%%%%%%%%%%%%%%%%%%%%%%%%%%%%%%%%%%%%%%%%%%%%%%%%%

%%%%%%%%%%%%%%%%%%%%%%%%%%%%%%%%%%%%%%%%%%%%%%%%%%%%%%%
%%% Appendix sections. ??????, ????
%%%%%%%%%%%%%%%%%%%%%%%%%%%%%%%%%%%%%%%%%%%%%%%%%%%%%%%

\begin{appendix}
	
\renewcommand{\thesection}{Appendix}

\section{}

In this appendix, we provide additional details regarding the parameters for GW190403\_051519. Table \ref{tableA3} presents the estimated parameters using Billy, while Table \ref{tableA4} presents the estimated parameters using pyRing.

\begin{table}[H]
	\renewcommand{\arraystretch}{1.4} 
	\tabcolsep 17pt
	\caption{Summary of the parameters for GW190403\_051519 that are estimated with Billy. $ \theta_1$ represents the tilt angle of the primary BH 's spin axis. $ \theta_2$ represents the tilt angle of the secondary BH 's spin axis. $ \Delta\Phi$ represents the azimuthal angle between the spin vectors of the two BHs. $\phi_{jl}$ represents the azimuthal angle between the total angular momentum vector and the orbital angular momentum vector. The median value and the range of the 90\% credible interval are provided. All masses are measured in the detector frame and are expressed in solar masses.}
	\label{tableA3}
	\begin{tabular*}{\columnwidth}{l c}
		\toprule
		Parameter & Value \\
		\hline
		The mass of the primary BH & $182.81_{-76.86}^{+40.47}$  \\
		The mass of the secondary BH & $33.79_{-11.51}^{+55.00}$  \\
		The spin of the primary BH & $0.83_{-0.71}^{+0.14}$  \\
		The spin of the secondary BH & $0.51_{-0.45}^{+0.43}$  \\
		$ \theta_1$ & $0.70_{-0.50}^{+1.75}$ \\
		$ \theta_2$ & $1.52_{-1.08}^{+1.12}$  \\
		$ \Delta\Phi$ & $3.02_{-2.70}^{+2.95}$  \\
		$\phi_{jl}$ & $3.31_{-2.93}^{+2.54}$  \\
		Declination  & $-0.56_{-0.47}^{+1.32}$  \\
		Right ascension & $2.56_{-0.94}^{+2.82}$ \\
		The binary's orbital inclination & $1.62_{-1.31}^{+1.25}$ \\
		The binary's polarization & $1.58_{-1.43}^{+1.40}$ \\
		The phase of coalescence & $3.40_{-2.99}^{+2.53}$\\
		\bottomrule
	\end{tabular*}
\end{table}

\begin{table}[H]
	\renewcommand{\arraystretch}{1.4} 
	\tabcolsep 18.5pt
	\centering
	\caption{Summary of the parameters for GW190403\_051519 that are estimated with pyRing. The median value and the range of the 90\% credible interval are provided. The mass of the remnant BH is measured in the detector frame and is expressed in solar masses.}
	\label{tableA4}
	\begin{tabular*}{\columnwidth}{l c}
		\toprule
		Parameter & Value \\
		\hline
		Polarization & $1.54_{-1.38}^{+1.44}$  \\
		The mass of the remnant BH & $206.63_{-48.73}^{+73.01}$  \\
		The spin of the remnant BH & $0.84_{-0.21}^{+0.14}$  \\
		Amplitude\_1 & $2.47_{-2.23}^{+8.03}$  \\
		Amplitude\_2 & $1.12_{-1.01}^{+3.34}$  \\
		Phase\_1 & $3.17_{-2.84}^{+2.78}$ \\
		Phase\_2 & $3.22_{-2.90}^{+2.78}$ \\
		\bottomrule
	\end{tabular*}
\end{table}

\end{appendix}

\end{multicols}

\begin{thebibliography}{99}
	
\bibitem{Penrose:1964wq}R. Penrose, \emph{Gravitational collapse and space-time singularities}, Phys. Rev. Lett. \textbf{14}, 57 (1965).

\bibitem{Penrose:1969pc}R. Penrose, \emph{Gravitational collapse: The role of general relativity}, Riv. Nuovo Cim. \textbf{1}, 252 (1969).

\bibitem{Hawking:1976ra} S. W. Hawking, \emph{Breakdown of Predictability in Gravitational Collapse}, Phys. Rev. D \textbf{14}, 2460 (1976).

\bibitem{wald2001thermodynamics}R. M. Wald, \emph{The thermodynamics of black holes}, Living Rev. Relativity \textbf{4}, (2001).

\bibitem{newwindow}X. Fan, \emph{The detection of gravitational waves and the new era of multi-messenger astronomy}, Sci. China Phys. Mech. Astron. \textbf{59}, 640001 (2016). 

\bibitem{astronomy}D. Blair, L. Ju, C. Zhao \textit{et al}. \emph{Gravitational wave astronomy: the current status}, Sci. China Phys. Mech. Astron. \textbf{58}, 120402 (2015)

\bibitem{abbott2016observation}R. Abbott \textit{et al}. (LIGO Scientific and Virgo Collaboration), \emph{Observation of gravitational waves from a binary black hole merger}, Phys. Rev. Lett. \textbf{116}, 061102 (2016).

\bibitem{review}R.-G. Cai, Z. Cao, Z.-K. Guo, S.-J. Wang, and T. Yang, \emph{The gravitational-wave physics}, Natl. Sci. Rev. \textbf{4}, 687 (2017)

\bibitem{review2}L. Bian, R.-G. Cai, S. Cao, Z. Cao, H. Gao, Z.-K. Guo, K. Lee, D. Li, J. Liu, Y. Lu, S. Pi, J.-M. Wang, S.-J. Wang, Y. Wang, T. Yang, X.-Y. Yang, S.
Yu, and X. Zhang, \emph{The Gravitational-wave physics II: Progress}, Sci. China-Phys. Mech. Astron. 64, 120401 (2021)

\bibitem{isi2021testing}M. Isi, W. M. Farr, M. Giesler, M. A. Scheel, and S. A. Teukolsky, \emph{Testing the black-hole area law with GW150914}, Phys. Rev. Lett. \textbf{127}, 011103 (2021).

\bibitem{isi2019testing}M. Isi, M. Giesler, W. M. Farr, M. A. Scheel, and S. A. Teukolsky, \emph{Testing the no-hair theorem with GW150914}, Phys. Rev. Lett. \textbf{123}, 111102 (2019).

\bibitem{PhysRevD.7.2333}J. D. Bekenstein, \emph{Black Holes and Entropy}, Phys. Rev. D \textbf{7}, 2333 (1973).


\bibitem{PhysRevLett.30.71}L. Smarr, \emph{Mass Formula for Kerr Black Holes}, Phys. Rev. Lett. \textbf{30}, 71 (1973).

\bibitem{bardeen1973four}J. M. Bardeen, B. Carter, and S. W. Hawking, \emph{The four laws of black hole mechanics}, Commun. Math. Phys. \textbf{31}, 161 (1973).

\bibitem{wald1994quantum}R. M. Wald, \emph{Quantum field theory in curved spacetime and black hole thermodynamics}, University of Chicago press, Chicago (1994).

\bibitem{thermodynamicallystable}Q. Wu, S.-W. Wei, and T. Zhu, \emph{Are the black hole remnants produced from binary black hole mergers in GWTC-3 thermodynamically stable?}, arXiv:2202.09290.




\bibitem{cabero2018observational}M. Cabero, C. D. Capano, O. Fischer-Birnholtz, B. Krishnan, A. B. Nielsen, A. H. Nitz, and C. M. Biwer, \emph{Observational tests of the black hole area increase law}, Phys. Rev. D \textbf{97}, 124069 (2018).

\bibitem{scientific2016tests}R. Abbott \textit{et al}. (LIGO Scientific and Virgo Collaboration), \emph{Tests of general relativity with GW150914}, Phys. Rev. Lett. \textbf{116}, 221101 (2016).

%\bibitem{bilbydocument}https://lscsoft.docs.ligo.org/bilby/.

\bibitem{bilby} G. Ashton \textit{et al}, \emph{BILBY: A User-friendly Bayesian Inference Library for Gravitational-wave Astronomy}, Astrophys. J. Suppl. Ser. 241, 27 (2019).


%\bibitem{pyRingdocument}https://lscsoft.docs.ligo.org/pyring/.

\bibitem{pyRing}G. Carullo, W. D. Pozzo, and J. Veitch, \emph{Observational black hole spectroscopy: A time-domain multimode analysis of GW150914}, Phys. Rev. D \textbf{99}, 123029 (2019).

%\bibitem{samples1}https://zenodo.org/record/6513631/.

%\bibitem{samples2}https://zenodo.org/record/5546663/.



\bibitem{GW190403}R. Abbott \textit{et al}. (LIGO Scientific and Virgo Collaboration), \emph{GWTC-2.1: Deep Extended Catalog of Compact Binary Coalescences Observed by LIGO and Virgo During the First Half of the Third Observing Run}, arXiv:2108.01045v2.


\bibitem{IMRPhenomXPHM}G. Pratten, C. Garc\'{i}a-Quir\'{o}s, M. Colleoni, A. Ramos-Buades, H. Estell\'{e}s, M. Mateu-Lucena, R. Jaume, M. Haney, D. Keitel, J. E. Thompson, and S. Husa, \emph{Computationally efficient models for the dominant and subdominant harmonic modes of precessing binary black holes}, Phys. Rev. D \textbf{103}, 104056 (2021).

 

\bibitem{abbott2021gwtc}R. Abbott \textit{et al}. (LIGO Scientific and Virgo Collaboration), \emph{GWTC-3: compact binary coalescences observed by LIGO and Virgo during the second part of the third observing run}, arXiv:2111.03606.









\bibitem{abbott2021gwtc1}R. Abbott \textit{et al}. (LIGO Scientific and Virgo Collaboration), \emph{GWTC-2: compact binary coalescences observed by LIGO and Virgo during the first half of the third observing run}, Phys. Rev. X \textbf{11}, 021053 (2021).





\bibitem{martynov2016sensitivity}D. V. Martynov \textit{et al}., \emph{Sensitivity of the Advanced LIGO detectors at the beginning of gravitational wave astronomy}, Phys. Rev. D \textbf{93}, 112004 (2016).

\bibitem{bersanetti2021advanced}D. Bersanetti, B. Patricelli, O. J. Piccinni, F. Piergiovanni, F. Salemi, and V. Sequino,  \emph{Advanced virgo: Status of the detector, latest results and future prospects}, Universe \textbf{7}, 322 (2021).

\bibitem{kagra2019kagra}KAGRA collaboration, \emph{KAGRA: 2.5 generation interferometric gravitational wave detector}, Nat. Astron. \textbf{3}, 35 (2019).

\bibitem{shaddock2008space}D. A. Shaddock, \emph{Space-based gravitational wave detection with LISA}, Classical Quantum Gravity \textbf{25}, 114012 (2008).

\bibitem{hu2017taiji}W.-R. Hu, Y.-L. Wu, \emph{The Taiji Program in Space for grav- itational wave physics and the nature of gravity}. Natl. Sci.
Rev. \textbf{4}, 685 (2017).

\bibitem{ruan2020taiji}W.-H. Ruan, Z.-K. Guo, R.-G. Cai, and Y.-Z. Zhang, \emph{Taiji program: Gravitational-wave sources}, Int. J. Mod. Phys. A \textbf{35}, 2050075 (2020).

\bibitem{luo2020brief}Z. Luo, Z. Guo, G. Jin, Y. Wu, and W. Hu, \emph{A brief analysis to Taiji: Science and technology}, Results Phys. \textbf{16}, 102918 (2020).

\bibitem{luo2016tianqin}J. Luo \textit{et al}. \emph{TianQin: a space-borne gravitational wave detector}. Class. Quantum Grav. \textbf{33}, 035010 (2016).

\bibitem{Gong:2021gvw}Y. Gong, J. Luo, B. Wang, \emph{Concepts and status of Chinese space gravitational wave detection projects}, Nat. Astron. \textbf{5}, 881 (2021).

\bibitem{amaro2018relativistic}P. Amaro-Seoane, \emph{Relativistic dynamics and extreme mass ratio inspirals}, Living Rev. Relativity \textbf{21}, 4 (2018).

\bibitem{berry2013observing}C. P. L. Berry, J. R. Gair, \emph{Observing the Galaxy's massive black hole with gravitational wave bursts}, Mon. Notices Royal Astron. Soc. \textbf{429}, 589 (2013).

\bibitem{babak2017science}S. Babak, J. Gair, A. Sesana, E. Barausse, C. F. Sopuerta, C. P. Berry, E. Berti, P. Amaro-Seoane, A. Petiteau, and A. Klein, \emph{Science with the space-based interferometer LISA. V. Extreme mass-ratio inspirals}, Phys. Rev. D \textbf{95}, 103012 (2017).

\bibitem{berry2013expectations}C. Berry and J. Gair, \emph{Expectations for extreme-mass-ratio bursts from the Galactic Centre}, Mon. Notices Royal Astron. Soc. \textbf{435}, 3521 (2013).

%\bibitem{gwosc}https://www.gw-openscience.org/.

\end{thebibliography}
\end{document}